\documentclass[12pt]{article}
\pdfoutput=1
\usepackage{epsfig}
\usepackage{amsfonts}
\usepackage{amssymb}

\begin{document}

\begin{center}

\vspace{1cm}

{\bf \large Challenges of D=6 \ N=(1,1)  SYM Theory} \vspace{1.5cm}

{\bf \large L. V. Bork$^{1,4}$, D. I. Kazakov$^{1,2,3}$, D. E. Vlasenko$^{2,3}$}\vspace{0.5cm}

{\it $^1$Alikhanov Institute for Theoretical and Experimental Physics, Moscow, Russia\\
$^2$Bogoliubov Laboratory of Theoretical Physics, Joint
Institute for Nuclear Research, Dubna, Russia.\\
$^3$Moscow Institute of Physics and Technology, Dolgoprudny, Russia\\
$^4$The Center for Fundamental and Applied Research, All-Russian Research Institute of Automatics, Moscow, Russia}
\vspace{0.5cm}

\abstract{Maximally supersymmetric  Yang-Mills theories have several remarkable properties, 
among which are the cancellation of UV divergences, factorization of higher loop corrections and possible integrability. 
Much attention has been attracted to the $\mathcal{N}=4$ $D=4$ SYM theory. The $\mathcal{N}=(1,1$) $D=6$ SYM theory
possesses similar properties but is nonrenomalizable and serves as a toy model for supergravity.
We consider the on-shell  four point scattering amplitude  and analyze  its perturbative expansion within the spin-helicity 
and superspace formalism. The integrands of the resulting diagrams coincide with those of the $\mathcal{N}=4$ $D=4$ SYM and 
obey the dual conformal invariance. Contrary to 4 dimensions, no IR divergences on mass shell appear. We calculate 
analytically  the leading logarithmic asymptotics in all loops. Their summation leads to a Regge trajectory which is calculated  exactly. The leading powers  of $s$ are calculated up to six loops. Their summation is performed numerically and leads to a smooth function of $s$. The leading UV divergences are calculated up to 5 loops. The result suggests the geometrical progression which ends up in a finite expression.
This leads us to a radical point of view on nonrenormalizable theories.
}
\end{center}

Keywords: Amplitudes, extended supersymmetry, UV divergences, Regge behaviour.

\section{Introduction}

In the last decade there has been considerable activity on the calculation of the amplitudes in maximally supersymmetric Yang-Mills theories  (SYM)~\cite{BDS4point3loop_et_all,D=5SYM_Diverges_ZBernDixon} and maximally supersymmetric gravity~\cite{N=8SUGRA finiteness}.  Gauge and gravity SUSY theories  in $D=4$ such as the $\mathcal{N}=4$ SYM and $\mathcal{N}=8$ SUGRA are the most important examples. These theories are believed to possess several remarkable properties, among which are total or partial cancelation cancellation of UV divergences, factorization of higher loop corrections and possible integrability. The success of factorization leading to the BDS ansatz~\cite{BDS4point3loop_et_all} for the amplitudes in 
$D=4$ $\mathcal{N}=4$ SYM stimulated similar activity in other models and dimensions.  Many magnificent insights in the structure of amplitudes (the S-matrix) of gauge theories in various dimensions (for review see, for example,~\cite{Reviews_Ampl_General}) were obtained.
It was understood that the structure of the integrands in all these theories is the same and has an imprint of conformal and dual conformal invariance~\cite{D4_DualConformal_Invariance,D6_DualConformal_Invariance,ZBern_GenUnit_D=6Helicity}. As a result, the structure of the UV divergences is also similar, in particular, the boundary where the first divergences in SYM appear happens to be given by the universal formula~\cite{2loopN=4 and finitenes bound 07 Bern&Co,N=4finitenessBound,1loopFinitnesBoud}
\begin{equation}\label{div}
D=4+6/L,
\end{equation}
where $D$ is  the dimension and $L$ is the number of loops. The structure of the amplitudes (and divergences) in SUGRA was also found to be linked to the SYM~~\cite{Grav_and_Gauge}. This renewed attempts to check the finiteness of the $D=4$ $\mathcal{N}=8$ SUGRA~\cite{N=8SUGRA finiteness,SUGRA fin Vs Div}.

All this activity became possible with the development of new techniques: the spinor helicity and momentum twistor formalisms, different sets of recurrence relations for the tree level amplitudes, the unitarity  based methods for loop amplitudes and different realizations of the on-shell superspace technique for theories with supersymmetry \cite{Reviews_Ampl_General}. These techniques were generalized to  a space-time dimension greater than $D=4$~ \cite{SpinorHelisityForm_GeneralDimentions_Boels,
DonaldOConnel_AmplInD=6, SpinorHelisityForm_D=10Dimentions}.

In this note, we consider one of these theories, namely, the  $D=6$ $\mathcal{N}=(1,1)$  SYM. This is a maximal supersymmetric theory in $D=6$ dimensions, after additional
compactification on two-torus it is reduced to the $D=4$ $\mathcal{N}=4$ SYM. It can also be  considered as a 
special low energy limit (the effective actions on the 5-branes \cite{AmplitudesAndM5Branes}) of the string/M theory.
It is believed that this theory  is also exceptional;  at the same time, it is nonrenormalizable by power counting, the coupling constant has a dimension $-2$ in mass units like in $D=4$ gravity. Therefore, this theory serves as a toy model for quantum gravity.

Investigation of this theory which we performed within the spinor-helicity  and superfield formalism has led us to some far-reaching conclusions. We first present our calculations which we performed up to 5 and 6 loops and  then make some speculations concerning nonrenormalizable theories.

\section{Color decomposition,  spinor helicity and superfield formalism in $D=6$ $\mathcal{N}=(1,1)$ SYM}\label{_2}

The aim is to calculate the multipaticle amplitudes on mass shell. For this purpose, we first perform the color decomposition extracting the color ordered partial amplitude~\cite{Reviews_Ampl_General}
\begin{equation}
\mathcal{A}_n^{a_1\dots a_n}(p_1^{\lambda_1}\dots p_n^{\lambda_n})=\sum_{\sigma \in S_n/Z_n}Tr[\sigma(T^{a_1}\dots T^{a_n})]
A_n(\sigma(p_1^{\lambda_1}\dots p_n^{\lambda_n}))+\mathcal{O}(1/N_c).
\end{equation}
The color ordered amplitude $A_n$ is evaluated in the planar limit which corresponds to $N_c\to \infty$, $g^2_{YM}\to 0$ and $g^2_{YM}N_c$ - fixed.

The next step is to use the spinor helicity formalism and on-shell
methods~\cite{Reviews_Ampl_General}.
 Their advantage  is that one calculates explicitly the physical amplitude with external states of a given helicity without unphysical degrees of freedom, gauge fixing, ghosts, etc., the usual attributes of a gauge theory.  The description of the spinor helicity formalism can be found in~\cite{DonaldOConnel_AmplInD=6,Sigel_D=6Formalism}.
Applying it one can rewrite the on shell amplitudes in a compact form. For example,
using the six dimensional version of the BCFW recurrence relation the tree level 4 gluon color ordered amplitude $A_4$ can be written as
\begin{eqnarray}\label{4gluon}
  \mathcal{A}_4^{(0)}(1_{a\dot{a}}2_{b\dot{b}}3_{c\dot{c}}4_{d\dot{d}})=
  -ig_{YM}^2
  \frac{\langle
  1_a2_b3_c4_d\rangle[1_{\dot{a}}2_{\dot{b}}3_{\dot{c}}4_{\dot{d}}]}{st},
\end{eqnarray}
where $1,2,3$ and $4$ are external momenta, $\langle 1_a2_b3_c4_d \rangle\doteq\epsilon_{ABCD}\lambda_{1}^{Aa}\lambda_{2}^{Bb}\lambda_{3}^{Cc}\lambda_{4}^{Dd}$ and $[ 1_{\dot{a}}2_{\dot{b}}3_{\dot{c}}4_{\dot{d}} ]$ $\doteq\epsilon^{ABCD} \tilde{\lambda}_{A,1}^{\dot{a}}
  \tilde{\lambda}_{B,2}^{\dot{b}}
  \tilde{\lambda}_{C,3}^{\dot{c}}
  \tilde{\lambda}_{D,4}^{\dot{d}}$, $\lambda_{i}^{Aa}$ and 
$\tilde{\lambda}_{A,i}^{\dot{a}}$ being the  spinors associated with momenta 
$p_{i}^{AB}$ of $i$'th particle. $A,B=1,\ldots,4$  are the
fundamental representation of the $Spin(SO(5,1))\simeq SU(4)^{*}$
indices, $a=1,2$ and $\dot{a}=\dot{1},\dot{2}$ are the $D=6$ little group
$SO(4) \simeq SU(2)\times SU(2)$ indices.
Note that in $D=6$ for the massless states helicity is no longer conserved in contrast to the $D=4$ case.

The superfield formalism allows one to take into account the full strength of the 
$\mathcal{N}=(1,1)$ supersymmetry. The self-­consistent way of constructing the  superamplitude comprises
the harmonic superspace techniques developed in~\cite{Sigel_D=6Formalism}.  It results in the following form of  the color ordered n-­particle superamplitude:
\begin{eqnarray}
 A_n(\{\lambda^A_a,\tilde{\lambda}_{A}^{\dot{a}},
\eta_a,\overline{\eta}_{\dot{a}} \})=
\delta^6(p^{AB})\delta^4(q^A)\delta^4(\overline{q}_A)\mathcal{P}_n(\{\lambda^A_a,\tilde{\lambda}_{A}^{\dot{a}},
\eta_a,\overline{\eta}_{\dot{a}} \}),
\end{eqnarray}
\begin{eqnarray}\label{projected_supercharges_n_particle_state}
  p^{AB}=\sum_i^n\lambda^{Aa}_i\lambda^B_{a,i},~~q^{A}=\sum_i^n\lambda^{A,i}_{a}\eta^{a}_i,~~
  \overline{q}_A=\sum_i^n\tilde{\lambda}_{A,i}^{\dot{a}}\overline{\eta}_{\dot{a},i},
\end{eqnarray}
$\lambda^A_a,\tilde{\lambda}_{A}^{\dot{a}}$ and
$\eta^{a},\overline{\eta}_{\dot{a}}$ being the bosonic and fermionic
coordinates of $\mathcal{N}=(1,1)$ on-shell momentum superspace, and
$\mathcal{P}_n$ is a polynomial with respect to $\eta$ and
$\overline{\eta}$ of degree of $2n-8$. 

We further  concentrate on the four point amplitude. In this case,
the degree of Grassmannian polynomial $\mathcal{P}_4$ is 0; 
hence $\mathcal{P}_4$ is a function of bosonic variables only
\begin{eqnarray}\label{4-point_ampl_general_form}
  A_4(\{\lambda^A_a,\tilde{\lambda}_{A}^{\dot{a}},
\eta_a,\overline{\eta}_{\dot{a}} \})= \delta^6(p^{AB})  \delta^4(q^A)\delta^4(\overline{q}_A)
  \mathcal{P}_4(\{\lambda^A_a,\tilde{\lambda}_{A}^{\dot{a}}\}).
\end{eqnarray}
Comparing this expression with (\ref{4gluon})  one concludes that the 
tree level 4-point
superamplitude can be written in a very compact form:
\begin{eqnarray}\label{4_point_tree_superamplitude}
  A_4^{(0)}=-ig_{YM}^2 \delta^6(p^{AB})\frac{\delta^4(q^A)\delta^4(\overline{q}_A)}{st}.
\end{eqnarray}

What is essential, at any order of PT the amplitude is proportional to  the bosonic and fermionic $\delta$-function of reflecting the (super)momentum
conservation as in (\ref{4-point_ampl_general_form}). This means that the tree level amplitude always factorizes and we get a universal expression for the color ordered superamplitude with the radiative corrections:
\begin{equation}
A_4(s,t)= A_4^{(0)}(s,t)\left[1+ \mbox{loop corrections} \right].
\end{equation}

\begin{figure}[ht]
\begin{center}
\leavevmode
\includegraphics[width=0.9\textwidth]{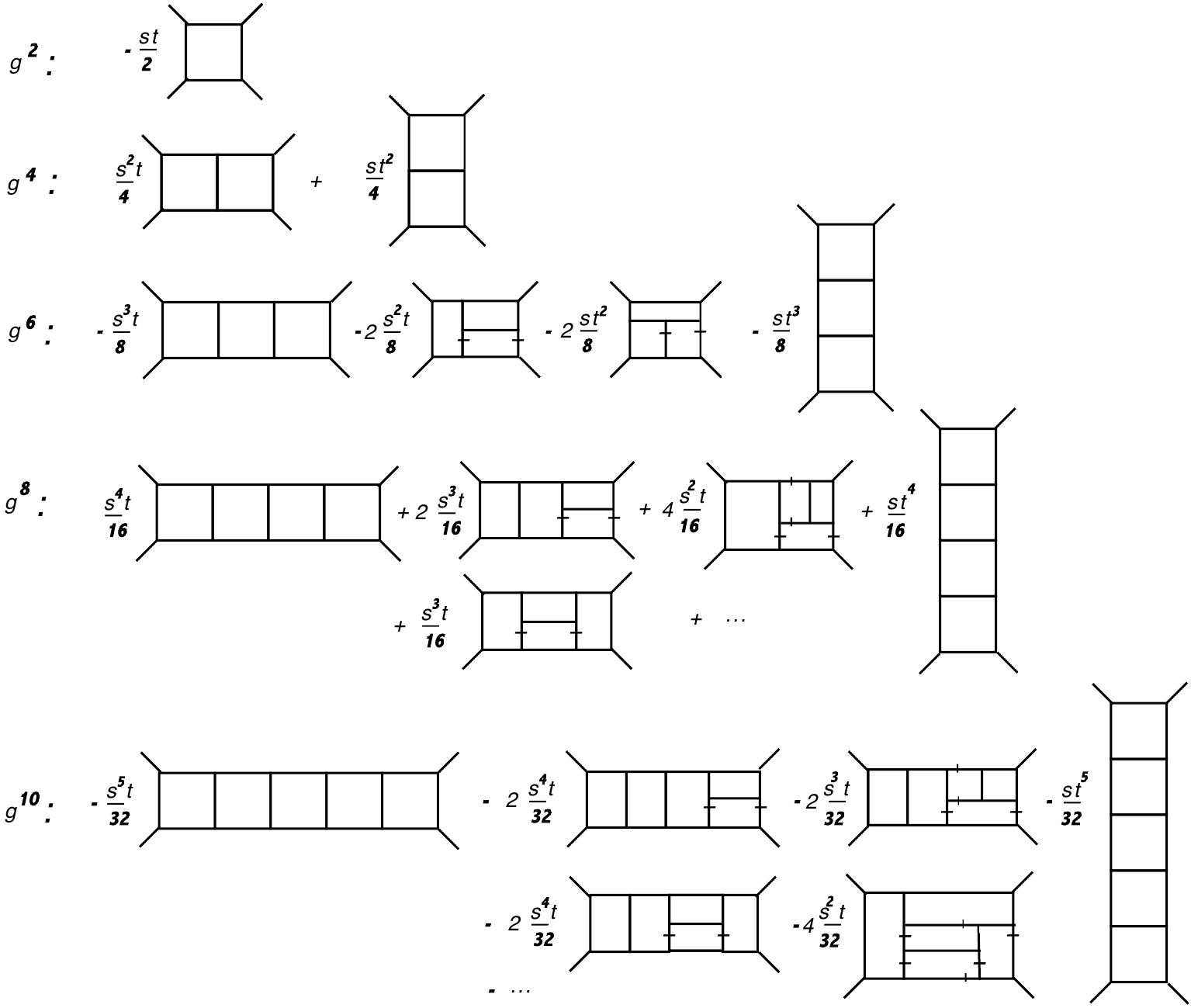}
\end{center}
\caption{The perturbative expansion for the scattering amplitude $A_4/A_4^{tree}$}
\label{exp}
\end{figure}

For the loop corrections one has expansion which due to a universal form of the integrands in any SYM theory coincides with the one in $D=4$ $\mathcal{N}=4$ SYM up to dimensional factors since in $D=6$ dimensions the coupling has a mass dimension equal to $-2$~\cite{ZBern_GenUnit_D=6Helicity}.  This is the consequence of the dual conformal invariance in momentum space~ \cite{D6_DualConformal_Invariance} equally valid in $D=4$ and in $D=6$. 
A remarkable property of this expansion is that all the bubble and triangle diagrams cancel and one is left with the sequence of scalar box diagrams shown in Fig.\ref{exp}.

\section{Perturbation Expansion for the Amplitudes}

Our task here is to calculate the radiative corrections to the four
point amplitude. In what follows we proceed loop by loop. The first question is: is there any kind of factorization similar to the  BDS formula? The answer is negative for general values of  Mandelstam variables $s$ and $t$
as it was shown in~\cite{BKV} where the two loop box diagram was calculated. While the one loop box gives the double logarithm of $s/t$, the two loop one contains the Polylog functions. In this situation, we concentrate on the Regge asymptotic behaviour when
$s\to \infty$ and $t<0$ is fixed.  Then,  all the integrals are expressed in terms of powers of $s$ and $t$ and $\log^{2n}(s/t)$. 

Since the coupling $g^2_{YM}$ in $D=6$  has dimension $-2$,  the expansion parameter is either $g^2_{YM}s$ or
$g^2_{YM}t$ and one can consider separately the series with the leading powers of $s$. In what follows we consider the infinite vertical series of  diagrams of Fig.\ref{exp} summing up the leading powers of $s$, the leading logarithms and the leading UV divergences. Note that the first UV divergence, in accordance with eq.(\ref{div}), is encountered in three loops.

\subsection{The Leading Logarithms}
In the Regge limit  the main contribution to the leading logs comes from
the vertical multiple boxes, the so-called ladder diagrams.
For the vertical n-loop ladder diagram $B_n(t,s)$, which is
UV and IR finite, the leading contribution  was found in~\cite{BKV} and takes the form
\begin{equation}\label{L-rung box assympt}
B_n(t,s)\simeq \frac 1s \frac{L^{2n}(x)}{n!(n+1)!}, \ \ \ \  L\equiv \log(s/t).
\end{equation}
Combined with the combinatorial factor $s (-\frac{t}{2})^n$  this leads to the series
or the leading logarithmical contributions ($L.L.$) to the amplitude 
\begin{equation}
\left.\frac{A_4}{A_4^{(0)}}\right|_{L.L.} = \sum_{n=0}^{\infty} \frac{(-g^2 t/2)^n L^{2n}(x)}{n!(n+1)!},
\ \ \ \mbox{where} \ \ g^2\equiv\frac{g_{YM}^2 N_c}{64\pi^3}.
\end{equation}
This series can be summed and represents the Bessel function of the imaginary argument
\begin{equation}\label{AmplitudeBessel}
\sum_{n=0}^{\infty} \frac{(-g^2 t/2)^n L^{2n}(x)}{n!(n+1)!}  = \frac{I_1(2 y)}{y}, \ \ \ y\equiv\sqrt{g^2|t|/2}\ 
L(x).
\end{equation}
In the Regge limit when $y\to\infty$ $I_1(2 y)\to \exp(2y)/(2\sqrt{\pi y})$
and one gets the Regge type behaviour
\footnote{Note that the tree amplitude
$A_4^{(0)}\sim s/t$  just like in four dimensions}
\begin{eqnarray}
\left. \frac{A_4}{A_4^{(0)}}\right|_{L.L.}\sim\left(\frac{s}{t}\right)^{\alpha(t)-1}
\end{eqnarray}
with
\begin{eqnarray}
  \alpha(t)=1+2\sqrt{g^2|t|/2}=1+\sqrt{\frac{g_{YM}^2 N_c|t|}{32\pi^3}}~.
\end{eqnarray}
One can
see that, as expected for the gauge theory, $\alpha(0)=1$.  Due to the square root,
the result resembles the one in the strong coupling regime and is exact in the limit $N_c\to\infty$.

\subsection{The Leading Powers}

Here we consider the leading powers (L.P.) of $s$ coming from the horizontal box diagrams $B_n(s,t)$ shown in Fig.\ref{exp}. These diagrams contain no UV divergences. Being the functions of $s$ and $t$ they do not contain the leading logs and have a finite limit when $t\to 0$. Hence, the $n$-loop horizontal box  can be represented in the form
\begin{equation}
B_n(s,t)=\frac{1}{s}\left(C_n+O(t/s)\right), \ \ n\geq 2,
\end{equation}
where the leading behaviour when $s\to\infty$ is given by the constant $C_n$.  
(The one loop case is somewhat special since it represents simultaneously the vertical  and horizontal boxes
contributing $\log^2(s/t)/2$  to the leading logs and $\pi^2/2$ to the leading powers.)
The first few constants $C_n$ up to 6 loops were calculated in~\cite{K}. The result is given in the Table:\vspace{0.5cm}
\begin{figure}[ht]
\begin{tabular}{|l|c|c|c|c|c|c|}
\hline 
Loops & 1 &2 & 3 & 4 & 5 & 6\\
\hline
Values & $\frac{\pi^2}{2}$ &  $\frac{\pi^2}{3}$& $ -\pi^2+\frac{31 \pi^6}{1890} $& & &\\
&&& $-8\zeta_3+4\zeta_3^2$ & & & \\ \hline
Numerics& 4.93 & 3.29& 2.06& 2.05& 2.42& 3.13\\
\hline
\end{tabular}\vspace{0.7cm}
\caption{The values of $C_n$ for the horizontal boxes (left) and the interpolation curve (right)}\label{values}
\end{figure}

\vspace{-4.9cm}\hspace*{10cm}
\includegraphics[width=0.3\textwidth]{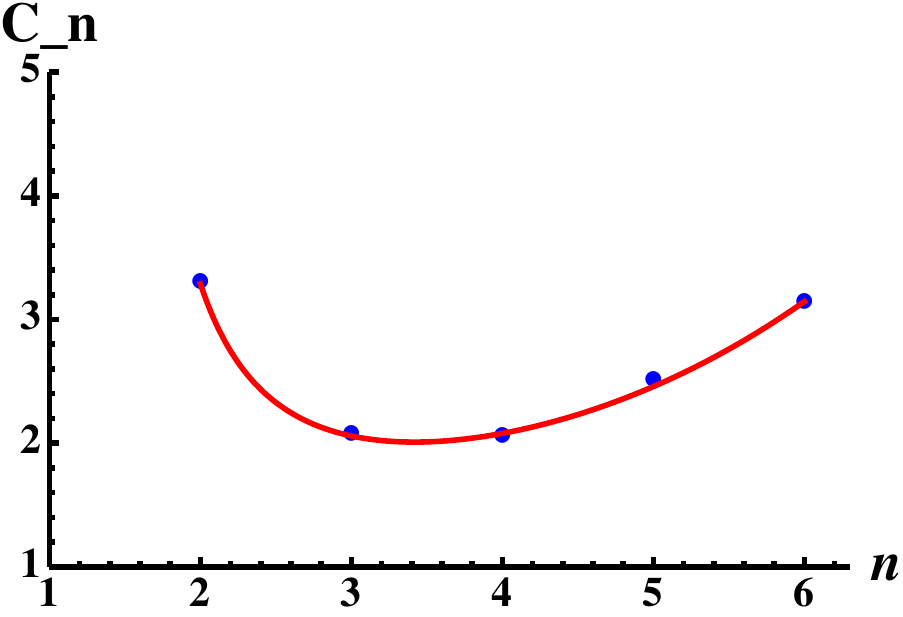}

\vspace{1.5cm}

One can see that $C_n$ do not have any simple iterative structure, so one is bound to use the numerical values.
The obtained numbers satisfy the following interpolation formula (see Fig.\ref{values}, right):
\begin{equation}
C_n\approx \frac{1.63^n}{1.31 n-1.80}\approx 0.76\frac{(\pi^2/6)^n}{n-4/3}, \ \ n\geq 2.
\end{equation}

Taking this formula one can sum the whole infinite series of Fig.\ref{exp} and get
\begin{eqnarray}
\left. \frac{A_4}{A_4^{(0)}}\right|_{L.P.}&\approx&-\frac{g^2t}{2}\left[\frac{\pi^2}{2}-\sum_{n=2}^\infty 0.76\frac{(-g^2s/2)^{n-1}(\pi^2/6)^n}{n-4/3}\right]\label{sumpower} \\
&&\hspace{-2cm}=-\frac{g^2t}{2}\frac{\pi^2}{2}\left[1-0.76\frac{g^2s\pi^2}{24} \phantom{}_2F_{1}(1,\frac 23,\frac 53,-\frac{g^2s\pi^2}{12})\right]
\approx -\frac{g^2t}{2}\frac{\pi^2}{2}\left[1-\left(\frac{g^2s }{2}\right)^{1/3}\right]. \nonumber
\end{eqnarray}
Note that the resulting function representing the infinite series of diagrams behaves differently from
the individual diagrams: while each individual diagram grows with $s$ as a power, the sum is a smooth function of $s$. The exact form of a function is not crucial here, eq.(\ref{sumpower}) demonstrates the general tendency.

\subsection{The Leading Divergences}

The UV divergences start from three loops. One has two ``tennis court" type diagrams shown in Fig.\ref{exp}. They both have the $1/\epsilon$ divergence but differ in the powers of $s$ and $t$. In what follows, we trace the leading divergences
of the form $1/\epsilon^n$ with the leading powers of $s$. 

In four loops one has $1/\epsilon^2$ terms, in five loops
$1/\epsilon^3$ terms and so on. These higher order divergences are linked to the lower order ones via the R-operation since all the divergences after subtraction of lower order subdivergences must have the local form. This is a general statement in any local field theory independently of the renormalizability. In principle, one can calculate the higher poles from the lowest one using the generalized reformalization group equation~\cite{Kazakov_nonerenorm_RG} though in practice this is not easy. 

The fact  that the higher order divergences are linked to the lower order
ones via the R-operation allows us to find the leading divergences iteratively without calculating the actual integrals. We performed the calculations up to 5 loops. In four loops one has 3 diagrams which have 3 loop divergent subgraphs, in five loops one has 10 diagrams which have  3 and 4 loop divergent subgraphs  and, consequently, the leading divergences. All these diagrams contain as a subgraph the ``tennis court" one. As a result, one has the following leading divergences with the leading power of $s$:
\begin{equation}
\begin{tabular}{|c|c|c|}
\hline
Loops &  Combinatorics & Divergence \\ \hline
3 & $(-g^2 s/2)^3\ \  2t/s$ & $1/6\epsilon$ \\ & & \\
4 & $(-g^2 s/2)^4\ \  2t/s $ & $1/36\epsilon^2$ \\  & & \\
5 & $(-g^2 s/2)^5\ \  2t/s $ & $1/216\epsilon^3$\\  \hline
\end{tabular}
\end{equation}
The form of the answer suggests the geometrical progression
\begin{equation}
\left. \frac{A_4}{A_4^{(0)}}\right|_{Leading\ Div.}= 2\frac ts \sum_{n=1}^\infty \left(-\frac{g^2s}{2}\right)^{n+2}\left( \frac{1}{6\epsilon}\right)^n= 
2\frac ts  \left(-\frac{g^2s}{2}\right)^{2}\frac{ \frac{-g^2s}{12\epsilon}}{1+ \frac{g^2s}{12\epsilon}}.
\end{equation}
When $\epsilon \to +0$ the sum has the finite limit
\begin{equation}
\left. \frac{A_4}{A_4^{(0)}}\right|_{Leading\ Div.} \to -2\frac ts  \left(-\frac{g^2s}{2}\right)^{2}= -\frac{g^4 s t}{2}.
\end{equation}

Thus, the sum of the infinite series is finite while each term diverges badly. The situation reminds the one in renormalizable theories where the leading logarithmic divergences sum up into the denominator. However, the crucial difference is that there we remove the divergences via the renormalization procedure absorbing them into the bare coupling, and here we cannot do it and take the limit $\epsilon\to +0$ without any renormalization. 
Contrary to the zero-charge case, the limit of the scattering amplitude is meaningful and leaves the finite part.
Presumably, the next-to-leading divergences behave similarly. 

\section{Discussion}

The obtained results of calculations lead us to the far-reaching conclusions. 

First of all, we see that  contrary to the renormalizable perturbation theory 
the finite number of terms does not give  the correct answer: The sum of the infinite series behaves differently from each individual term. This is true for both the leading powers and the leading logarithms. The summation of the whole infinite series of the leading logarithms gives the power law behaviour while the summation of the leading powers gives the smooth function.  It may well  be that the Regge behaviour obtained above is correct in the full theory.

Second, the usual perturbation theory  is badly divergent in each finite order while the whole series seems to be finite!
This is a remarkable property of the series  which we checked up to 5 loops for the leading divergences and leading powers. Needless to say that ma ore thorough check of this fact in the next-to-leading order and subleading powers would be  very  desirable.

If what we have conjectured is true, it might mean that
 in nonrenormalizable theories the finite number of PT terms has no meaning while the full theory exists.
 That would imply that  severe UV divergences  present in any given order of PT are actually artifacts of
 the weak coupling expansion. Since the model at hand is the toy model for gravity, one may try to apply the same arguments to quantum gravity. This would mean that one should not be confused by 
 nonrenormalizability of PT in quantum gravity. It may well be that the full theory is meaningful, PT is just not applicable here.

All above considerations bring us to the final conclusion that in order to understand the nonrenormalizable theories, one has to find an alternative description. On has to calculate the amplitude abandoning the usual PT. It should be some kind of a dual description similar to the AdS/CFT correspondence, reggeization of gluons~\cite{Lipatov}, description of observables in different terms, etc.  For example, it is known that the action of nonrenormalizable $D=3$ quantum gravity can be written in the form of a pure Chern-Simons theory~\cite{Witten_D3_Grav}, which is finite. The result of an alternative approach might be quite different from the PT one.

\section*{Acknowledgements}
 The authors are grateful to V.Smirnov for the evaluation of the 3 loop divergent coefficient. Financial support from RFBR grant \# 14-02-00494  and Ministry of Science and Education grant \# H\underline{III}-3810.2010.2 is acknowledged.

\end{document}